\documentclass[%
    aip,
    jcp,
    amsmath,
    amssymb,
    twocolumn,
    superscriptaddress,
    reprint]{revtex4-1}
\usepackage{dcolumn}
\usepackage{graphicx} 
\usepackage{bm}
\usepackage{longtable}
\usepackage[english]{babel}
\usepackage{color}

\begin{document}

\title{Configurational entropy of polar glass formers and the effect of electric field on glass transition} 
\author{Dmitry V.\ Matyushov}
\email{dmitrym@asu.edu}
\affiliation{Department of Physics and School of Molecular Sciences, 
         Arizona State University, PO Box 871504, Tempe, Arizona 85287}

\begin{abstract}
A model of low-temperature polar liquids is constructed that accounts for configurational heat capacity, entropy, and the effect of a strong electric field on the glass transition. The model is based on Pad{\'e}-truncated perturbation expansions of the liquid state theory. Depending on parameters, it accommodates an ideal glass transition of vanishing configurational entropy and its avoidance, with a square-root divergent enumeration function at the point of its termination.  A composite density-temperature parameter $\rho^\gamma/T$, often used to represent combined pressure and temperature data, follows from the model. The theory is in good agreement with experimental data for excess (over the crystal state) thermodynamics of molecular glass formers. We suggest that the Kauzmann entropy crisis might be a signature of vanishing configurational entropy of a subset of degrees of freedom, multipolar rotations in our model. This scenario has observable consequences: (i) a dynamical cross-over of the relaxation time and  (ii) the fragility index defined by the ratio of the excess heat capacity and excess entropy at the glass transition. The Kauzmann temperature of vanishing configurational entropy, and the corresponding glass transition temperature, shift upward when the electric field is applied. The temperature shift scales quadratically with the field strength.  
\end{abstract}

\maketitle 

\section{Introduction}
\label{sec:1}
Configurational entropy in statistical mechanics enumerates the number of states of a macroscopic system available at a given value of its energy.\cite{StillingerBook} It is defined through the density of states $\Omega(E)$ entering the canonical partition function 
\begin{equation}
  e^{-\beta F(\beta)} = \int \Omega (E) e^{-\beta E} dE .
\label{eq:1}    
\end{equation}
Here, $\beta =1/(k_\text{B}T)$ is the inverse temperature and $F(\beta)$ is the system free energy. From this relation, the configurational entropy is the logarithm of the density of states evaluated at the average energy of the system $\bar E$
\begin{equation}
S_c = \ln \left[ \Omega(\bar E)\right] .
\label{eq:2}  
\end{equation}
Here and below, the entropy is given in units of $k_\text{B}$ and $\bar E= \bar E(T)$ is a function of temperature at fixed volume/pressure. Correspondingly, $S_c=S_c(T)$ is a function of temperature at isochoric or isobaric conditions.  

The density of states is formally calculated by counting the number of states consistent with a given potential energy $E$
\begin{equation}
\Omega(E) = (\lambda^{3N} N!)^{-1}\int \delta (E- V_N) e^{-\beta V_N} d\mathbf{r}^N,
\end{equation}
where $V_N$ is the potential energy of the system of $N$ particles and $\lambda$ is the thermal de Broglie wavelength.\cite{Sciortino:05} For an ideal gas, $V_N=0$ and one gets the corresponding density of states $\Omega(E)=\delta(E)V^N/(\lambda^{3N}N!)$.

Mathematically, Eq.\ \eqref{eq:1} is the Laplace integral in the energy variable. Therefore, the density of states follows from the inverse Laplace transform in the variable $\beta$.\cite{Freed:03} The calculation of such an inverse transform is performed here  following an earlier publication.\cite{DMpre:07} This approach is applied to the free energy of a polar liquid obtained from a Pad\'e-truncated perturbation expansion in the angular (multipolar) potential.\cite{Gubbins:84} The result is a non-Gaussian enumeration function, $\sigma = N^{-1}\ln[\Omega]$, applied here to analyze experimental data for supercooled molecular glass formers and to develop a model of the effect of electric field on glass transition. 

Configurational entropy has played a significant role in the theory of glass transition,\cite{Stillinger:2013tl} which is a kinetic phenomenon of ergodicity breaking under the kinetic slowing down. The connection between kinetics and thermodynamics is sought by the Adam-Gibbs (AG) theory,\cite{Adam:65} which maintains that slowing dynamics has its thermodynamic origin in a decreasing number of configurations which a low-temperature liquid can potentially explore. The mathematical link between the increasing time of structural $\alpha$-relaxation $\tau(T)$ and the configurational entropy is through the AG relation, $\ln[\tau(T)/\tau_0] \propto [TS_c(T)]^{-1}$ ($\tau_0\simeq 10^{-14}-10^{-13}$ s is the characteristic vibrational time). From this equation, the drop of $S_c(T)$ to zero, when the ideal glass state with a single configuration is achieved, signifies the divergence of the relaxation time beyond any time-scale attainable by measurements, $\tau(T)\to\infty$. 

The AG theory has enjoyed significant support from empirical evidence.\cite{Angell:95} In particular, the extrapolated temperature of vanishing entropy, the Kauzmann temperature $T_K$, is often found to be close to the extrapolated temperature at which the relaxation time formally diverges.\cite{Richert:98} The fitting of the relaxation time is typically done with the Vogel-Fulcher-Tammann (VFT) relation $\ln[\tau/\tau_0] \propto (T-T_0)^{-1}$, from which the divergence temperature $T_0$ is found to be close to $T_K$. The VFT equation is based on empirical evidence and the dynamical divergence might be an artifact of the mathematics.\cite{Hecksher:2008co} However, a number of theories, most notably the random first-order transition theory (RFOT), support a direct link between slowing dynamics and decreasing configurational entropy.\cite{Lubchenko:2007cc} Importantly, $T_0=T_K$ is explicitly assumed in the RFOT to connect the configurational thermodynamics to relaxation. Despite its importance, there are very few reliable mathematical functionalities that can be used to model the configurational entropy of condensed materials.\cite{Derrida:80,Moynihan:00,Shell:04,DMjcp1:07} If the relaxation dynamics and configurational thermodynamics are indeed related,\cite{Ito:99,Martinez:01,Wang:2006cb} it would be beneficial to develop exactly solvable models for the configurational entropy and to explore thermodynamic forces alternative to broadly used temperature and pressure to consistently perturb both the dynamics and thermodynamics.   

Electric field traditionally employed in dielectric spectroscopy has recently emerged as an additional thermodynamic force to affect both the statistics and dynamics of polar liquids. Linear dielectric spectroscopy has been widely used to study dynamical properties of equilibrium and super-cooled polar liquids.\cite{Lunkenheimer:2010dz,Richert:2014wa} However, linear response does not affect the structure of the material and, therefore, does not modify either structural dynamics or configurational entropy. Altering structure requires electric fields sufficiently strong to produce a measurable non-linear dielectric response.\cite{Richert:2014wa} Along this line of thought, Johari has recently suggested to use strong electric fields to further test the significance of configurational entropy in the glass transition.\cite{Johari:2013hq}    

Johari's suggestion assumes that the configurational entropy of a bulk material is modified by the electric field, and this modification can be estimated by adding the thermodynamic entropy of material's polarization\cite{Landau8} to the entropy of unpolarized material
\begin{equation}
S_c(\mathcal{E},T) = S_c(T) + (\mathcal{E}^2/8\pi) (\partial \epsilon/\partial T)_V.
\label{eq:3}
\end{equation}
Here, $\mathcal{E}$ is the macroscopic (Maxwell) field in the sample with the dielectric constant $\epsilon(T)$. 

The idea that a thermodynamic entropy can be simply added to the configurational entropy is highly questionable to begin with. General arguments\cite{StillingerBook} and specific calculations\cite{Dudowicz:2006hi,DMjcp1:07} suggest that configurational entropy enumerates the number of states available to elementary excitations in the liquid induced by thermal agitation. Altering the configurational entropy has to change the spectrum of these, local or collective, excitations repopulating some of them relative to the others. Merely adding an entropy derived on thermodynamic grounds assuming linear response does not seem to accomplish this goal. However, supporting these generic arguments requires a specific landscape model, and this is what this article is set out to accomplish.  

We apply here the general perturbation theory of polar liquids\cite{Gubbins:84} to derive an exact analytical form for the enumeration function yielding the configurational entropy. The landscape model is non-Gaussian, and it requires three independent parameters to produce the temperature-dependent configurational entropy and heat capacity. In order to test the performance of the model, it is used to fit experimental data for the excess (relative to the crystal) entropies and heat capacities of molecular glass formers. The perturbation expansion is then extended to the case of a liquid polarized by a uniform external field. This extension is particularly productive in the context of thermodynamics of polar liquids since the coupling of dipoles to the external field adds to the Hamiltonian of anisotropic interactions of the liquid multipoles and thus enters the same perturbation formalism in terms of anisotropic, orientation-dependent interactions. The alteration of the configurational thermodynamics by the external field is therefore expressed in terms of the same model parameters and permits an additional test of the model by experiment. It also provides the experimental input helping to parametrize the model.  

Independently from the specifics of the model and as anticipated from general arguments, adding the free energy of polarizing the dielectric, $F_E=E_E-TS_E$, to the free energy of non-polarized polar liquid does not modify the energy landscape, but only shifts the relevant energies and the enumeration function (see below). However, the modification of the perturbation expansion by the external field does alter the enumeration function beyond a simple shift and changes the configurational thermodynamics.

\section{Non-Gaussian Landscape}
\label{sec:3}
We consider here a liquid of polar molecules interacting by nonpolar, Lennard-Jones (LJ) type interactions and by multipolar interactions. One can, therefore, separate the interaction potential into a radial (spherically-symmetric) part $H_0$ and an angular part $H_a$ depending on molecular orientations. One possible way, adopted here, to proceed with calculating the thermodynamic properties of such a liquid is to apply the perturbation expansion in terms of the angular interaction energy $H_a$ while adopting the isotropic distribution functions obtained with $H_0$ as reference (zero-order perturbation).\cite{Gubbins:84} The free energy of the liquid 
\begin{equation}
F = F_0 +\Delta F = F_0 - F_2 + F_3+\dots
\label{eq:10}  
\end{equation}
becomes a sum of the reference, non-polar part $F_0$ and a perturbation expansion for the polar part $\Delta F$. The expansion terms can be directly calculated:\cite{Gubbins:84} $F_2=(\beta/2)\langle H_a^2\rangle$ and $F_3=(\beta^2/6)\langle H_a^3\rangle$.  
   
The expansion in Eq.\ \eqref{eq:10} is typically difficult to calculate beyond $F_3$ and truncation is required. A Pad\'e form to truncate the perturbation series was suggested by Stell and co-workers.\cite{Stell:77,Gubbins:84} It replaces $\Delta F$ with the following form
\begin{equation}
\Delta F = - \frac{F_2}{1+F_3/F_2} = - Ne_a \frac{\beta^{*2}}{1+\beta^*},  
\label{eq:11}
\end{equation}
which is exact for the first two expansion terms and generates a sign-alternating infinite series, as expected. 
In Eq.\ \eqref{eq:11}, $N$ is the number of liquid particles and we have introduced the reduced inverse temperature 
\begin{equation}
\beta^* = T'/T = (\beta/3) \langle H_a^3\rangle /\langle H_a^2 \rangle .
\label{eq:12}  
\end{equation}
Further, since the system free energy and the expansion terms $F_{2,3}$ are extensive, the parameter 
\begin{equation}
e_a=(9/2N)\langle H_a^2\rangle^3/\langle H_a^3\rangle^2
\label{eq:12-1}  
\end{equation}
is intensive. In practical calculations, it is given by a combination of perturbation integrals arising from the perturbation expansion with the reference distribution functions of the nonpolar liquid.\cite{Larsen:77} We, however, do not pursue this direction here and limit ourselves to considering a general functionality of the density of states and the configurational entropy as produced by Pad\'e-truncated perturbation formalisms.\cite{Gubbins:84} It suffices therefore to note that $e_a$, as expressed through the corresponding perturbation integrals, is a function of density, which is held constant when the inverse Laplace transform over $\beta^*$ is performed below. This parameter is therefore a constant for a given liquid held at a constant density. It has the meaning of the overall energy of multipolar stabilization when the liquid is cooled down (see below).

The free energy $F_0=E_0-TS_0$ is composed of the energy $E_0$ of LJ attractions and the free energy of packing the repulsive cores of the molecules. The former is mostly temperature independent and does not contribute a significant entropy component.\cite{WCA} The latter is mostly entropic and can be approximated by the entropy of packing the molecular repulsive cores. Each of these components, $E_0$ and $S_0$, can to a good approximation be viewed as temperature independent at constant density. Equations \eqref{eq:10} and \eqref{eq:11} can now by used in Eq.\ \eqref{eq:1} to produce the inverse Laplace transform in the variable $\beta^*$
\begin{equation}
\Omega(e) = e^{S_0}\int_{c-i\infty}^{c+i\infty} \frac{d\beta^*}{2\pi i} \exp\left[
N\beta^*e + N\beta^{*2}e_a/(1+\beta^*)\right],
\label{eq:13}  
\end{equation}
where $e= \beta'(E-E_0)/N$ and $\beta'=1/(k_\text{B}T')$ [Eq.\ \eqref{eq:12}]. Following briefly the steps of Ref.\ \onlinecite{DMpre:07}, one can expand the exponent of the second term in the brackets, followed by the residue calculus. The result is a closed-form expression
\begin{equation}
\Omega(e)=\frac{e^{S_0}}{\sqrt{1+e/e_a}}e^{-N(e+2e_a)}I_1(2N\sqrt{e_a(e_a+e)}),
\label{eq:13-1}  
\end{equation}
where $I_1(x)$ is the modified Bessel function. This equation can be asymptotically expanded in the thermodynamic limit $N\to\infty$, with the resulting enumeration function\cite{StillingerBook} $\sigma(e)=N^{-1}\ln\left[\Omega(e)\right]$ in the form
\begin{equation}
\sigma(e) = \sigma_\infty - \left(\sqrt{e+e_a} - \sqrt{e_a}  \right)^2.
\label{eq:14}  
\end{equation}
Here, $\sigma_\infty=S_0/N$ specifies the top of the energy landscape enumerating the number of accessible configurations per liquid molecule in the nonpolar reference fluid with $e\to 0$ and $E\to E_0$.  

The energy landscape is clearly non-Gaussian, but it contains the parabola of the Gaussian random energy model\cite{Derrida:80,Heuer:00} in the limit $e_a\gg e$. The expansion of the square root in powers of $e/e_a$ produces the Gaussian form
\begin{equation}
  \sigma(e) = \sigma_\infty - e^2/(4e_a). 
\label{eq:15}  
\end{equation}

One can next calculate the average energy $\bar e$ from the first derivative $d\sigma/de=T'/T$ and the heat capacity from the second derivative of the enumeration function $d^2\sigma/de^2=-c_c^{-1}(T'/T)^2$, where the constant volume configurational heat capacity $c_c$ per molecule of the liquid is in units of $k_\text{B}$.  This calculation yields for these functions
\begin{equation}
\begin{split}
  \bar e & = -e_a \frac{1+ 2(T/T')}{(1+ T/T')^2}, \\
  c_c & = 2e_a \frac{T/T'}{(1+ T/T')^3} . 
\end{split}
\label{eq:16}  
\end{equation}
The energy $\bar e=-e_a$, achieved at $T=0$, establishes the overall drop of the energy of multipolar interactions upon cooling the liquid from the level $\bar e=0$ at $T\to\infty$, when only LJ interactions contribute to the internal energy.  

By substituting the average energy $\bar e$ into the enumeration function, one arrives at the configurational entropy
\begin{equation}
s_c = S_c/N = \sigma_\infty\left[1 - \frac{\tau}{(1+ T/T')^2}  \right],
\label{eq:17}  
\end{equation}
where 
\begin{equation}
\tau = e_a/\sigma_\infty
\end{equation}
is an effective temperature. Equations \eqref{eq:16} and \eqref{eq:17} also lead to a simple relation between the configurational entropy and configurational heat capacity
\begin{equation}
 s_c(T) = \sigma_\infty - \left[\sqrt{e_a}c_c(T)(T'/2T)\right]^{2/3} .  
 \label{eq:18}
\end{equation}
Overall, the configurational thermodynamics is defined by three parameters: $e_a$ and $T'$ are required for the average energy and heat capacity and an additional parameter, the high-temperature entropy $\sigma_\infty$, is required for the configurational entropy. 

\begin{figure}
\includegraphics*[clip=true,trim= 0cm 1.5cm 0cm 0cm,width=6.5cm]{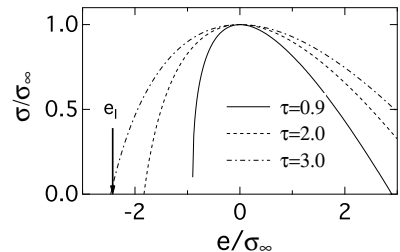}  
\caption{{\label{fig:1}}Enumeration function from Eq.\ \eqref{eq:14} at different values of $\tau$ indicated in the plot; the vertical arrow shows the energy of the ideal glass state $e_I$ at which $\sigma(e_I)=0$. 
}
\end{figure}

In order to appreciate the distinction between the non-Gaussian energy landscape presented here and the standard random energy model [Eq.\ \eqref{eq:15}] it is useful first to turn to the enumeration function. Figure \ref{fig:1} shows representative curves of $\sigma(e)/\sigma_\infty$ plotted against $e/\sigma_\infty$ at different values of the effective temperature $\tau=e_a/\sigma_\infty$. The first point to address is the ability of the system to achieve the state of ideal glass, when it runs out of configurations and $\sigma(e_I)=0$.\cite{Sciortino:99,Stillinger:2013tl} This limit is achieved only when $\tau\ge 1$, when the ideal glass energy is 
\begin{equation}
  e_{I}/\sigma_\infty=1-2\sqrt{\tau}.
\label{eq:18}  
\end{equation}
This energy is achieved at the Kauzmann temperature 
\begin{equation}
  T_K/T' = \sqrt{\tau}-1 .
  \label{eq:18-1}
\end{equation}

If $\tau<1$, the enumeration function ends with a residual entropy and an infinite derivative at its lowest point $e=-e_a$ (solid curve in Fig.\ \ref{fig:1}). This state is reached only at $T=0$ and the ideal glass is avoided. This scenario is similar to the avoided ideal glass suggested by Stillinger,\cite{Stillinger:88,Stillinger:2013tl} except that the divergence of the derivative is inverse square root, instead of the logarithmic divergence in Stillinger's analysis. No point of divergence of $\sigma'(e)$ appears in the Gaussian landscape model.

\begin{figure}
\includegraphics*[clip=true,trim= 0cm 1.5cm 0cm 0cm,width=6cm]{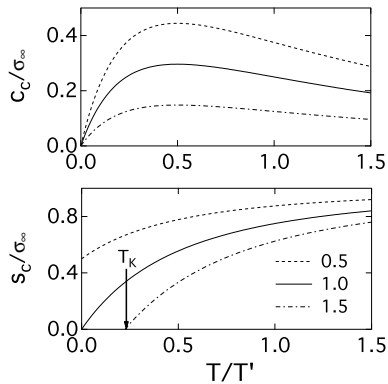} 
\caption{{\label{fig:2}} Heat capacity (upper panel) and configurational entropy (lower panel) calculated in the present models with the values of the parameter $\tau=e_a/\sigma_\infty$ shown in the plot. The vertical arrow in the lower panel indicates the Kauzmann temperature $T_K$. 
} 
\end{figure}

Figure \ref{fig:2} shows representative temperature plots for $s_c(T)$ and $c_c(T)$. As is already clear from Fig.\ \ref{fig:1}, the parameter $\tau$ controls the ability of the system to reach the state of the ideal glass at a positive temperature. At $\tau<1$, the drop of the configurational entropy ends at a positive residual value at $T=0$, while the Kauzmann temperature $T_K>0$, $s_c(T_K)=0$ is reached at $\tau>1$. Below we apply this landscape model to experimental thermodynamic and relaxation data of molecular glass formers.

\begin{table}
  \caption{Parameters of simultaneous fitting of experimental excess heat capacity and excess entropy\cite{Moynihan:00,Tatsumi:2012iy} to the landscape model (Fig.\ \ref{fig:3}). $T'$, $\sigma_\infty$, and $e_a$ fully define the energy landscape model. }
  \label{tab:1}
  \begin{ruledtabular}
  \begin{tabular}{lccccc}
Liquid & $T'$, K & $\sigma_\infty$ & $e_a$$^a$ & $T_K$, K$^b$ & $T_0$, K$^c$\\
\hline
OTP$^d$  & 142 & 14 & 80 & 203 & 202\\
Toluene  & 72 & 8.4 & 48 & 100 & 97\\
MTHF$^e$ & 56 & 9.9 & 50 & 69 & 70\\
Salol    & 131 & 14 & 76 & 174 & 175\\
1-butene$^f$   & 23  & 8.1 & 80 & 50 & \\
\end{tabular}
\end{ruledtabular}  
$^a$ $e_a$ is converted to the energy units my multiplying with $k_\text{B}T'$. $^b$calculated from Eq.\ \eqref{eq:18-1}. $^c$from Ref.\ \onlinecite{Richert:98}. $^d$\textit{o}-terphenyl. $^e$2-methyltetrahydrofuran. $^f$data from Ref.\ \onlinecite{Tatsumi:2012iy}.  
\end{table}

\section{Comparison to experiment}
\begin{figure}
\includegraphics*[clip=true,trim= 0cm 1.5cm 0cm 0cm,width=6cm]{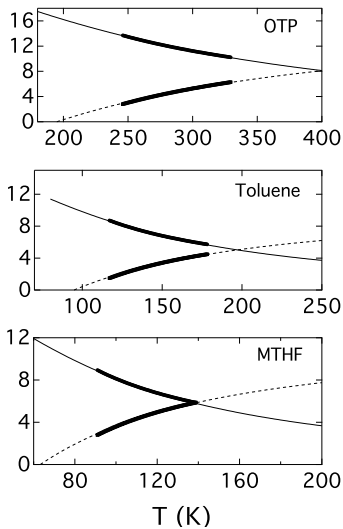} 
\caption{{\label{fig:3}} Heat capacities (upper curves) and excess entropies (lower curves) from experiment (points) expressed in units of $k_\text{B}$. The lines are fits to Eqs.\ \eqref{eq:16} and \eqref{eq:17} for the configurational heat capacity (solid curves) and the configurational entropy (dashed curves). The fitting parameters are listed in Table \ref{tab:1}.  
} 
\end{figure}

Configurational entropies are not available experimentally and excess entropy $s_\text{ex}$ of the supercooled liquid over its crystalline state is often used instead, $s_\text{ex}(T)\simeq s_c(T)$.\cite{Moynihan:00,Stevenson:05,DMjcp1:07}  Correspondingly, one puts $c_\text{ex}(T)\simeq c_c(T)$ for the heat capacity. This assignment assumes that the vibrational density of states does not alter between the crystal and supercooled liquid and thus the vibrational entropy and heat capacity cancel out in the difference. We have applied the functionality derived above to simultaneously fit $s_\text{ex}(T)$ and $c_\text{ex}(T)$ for four common molecular glass formers.\cite{Moynihan:00,Tatsumi:2012iy} The quality of the fit is shown in Fig.\ \ref{fig:3} and the fitting parameters are listed in Table \ref{tab:1}.   It is clear that all liquids in Table \ref{tab:1} fall in the regime of $\tau>1$ with $T_K>0$.

\begin{figure}
\includegraphics*[clip=true,trim= 0cm 1.5cm 0cm 0cm,width=6cm]{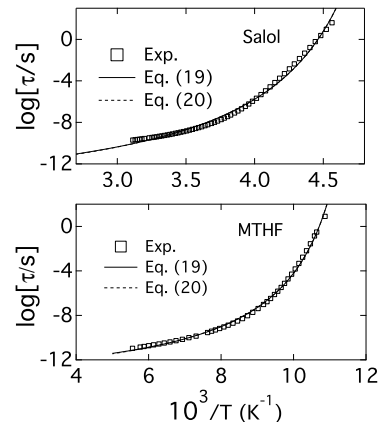} 
\caption{Dielectric relaxation time of salol\cite{Stickel:1995du} and MTHF\cite{Richert:98} (points) compared to fits to Eqs.\ \eqref{eq:20} (solid line) and \eqref{eq:21} (dashed line) assuming $c_c(T)\simeq c_\text{ex}(T)$. The two sets of lines are practically indistinguishable on the scale of the plots. The VFT equation, not displayed, provides comparable fit quality. 
}
\label{fig:4}  
\end{figure}

The present model suggests the following form of the AG relation for the relaxation time $\tau$
\begin{equation}
  \label{eq:20}
  \ln(\tau/\tau_0) = \frac{A\tilde T}{T-(Tc_c(T)^2\tilde T^2)^{1/3}}.
\end{equation}
Here, $A$ and $\tilde T$ are fitting parameters, the former is dimensionless and the latter is an effective temperature. Equation \eqref{eq:20} carries functionality similar to the one derived in the excitation model of the configurational entropy\cite{DMjcp1:07}
\begin{equation}
  \label{eq:21}
  \ln(\tau/\tau_0) = \frac{A\tilde T}{T-\tilde T c_c(T) }.
\end{equation}
Equation \eqref{eq:21} is consistent with experimental data\cite{DMjcp1:07} and  the two analytical forms are in most cases indistinguishable by experiment. This is illustrated in Fig.\ \ref{fig:4}, where they are used to fit the experimental dielectric relaxation times of salol\cite{Stickel:1995du} and  2-methyltetrahydrofuran (MTHF).\cite{Richert:98}  The fit quality is consistently worse for salol, but both formulas, Eqs.\ \eqref{eq:20} and \eqref{eq:21}, produce very close fits. They are also close to the corresponding VFT fits (not shown in Fig.\ \ref{fig:4}).

One can arrive at a slightly modified VFT equation from the present formalism by using $s_c(T)$ from Eq.\ \eqref{eq:17} in the AG equation
\begin{equation}
  \ln(\tau/\tau_0) = \frac{A f(T)}{T-T'(\sqrt{\tau}-1)} ,
  \label{eq:22}
\end{equation}
where $f(T)$ is a weak function of temperature. As mentioned above, at $\tau>1$ one gets the dynamical divergence of the VFT type, which disappears at $0<\tau<1$. The present model thus allows both $T_0>0$ and $T_0<0$ in the VFT equation.

Relaxation times measured at different temperatures and pressures can often be superimposed on a single master curve by considering the combined density-temperature thermodynamic variable $\rho^\gamma/T$,\cite{Gundermann:2011bw} where $\gamma$ is a material constant found to vary in a wide range, $0.1<\gamma<9$, between different glass formers.\cite{Roland:2005ga} The present model offers a potential route to this empirical rule, although the magnitude of $\gamma$ seems to be difficult to establish, in agreement with observations. 

It is clear from the derivation that the effective temperature entering the model is $T/T'$, where $T'$ is given by Eq.\ \eqref{eq:12}. For the perturbation expansions in terms of dipole-dipole molecular interactions $T'$ becomes\cite{Larsen:77}
\begin{equation}
 k_\text{B}T' = \frac{m^2}{9\sigma_s^3} \rho^* \frac{I_{TD}}{I_6} ,
\label{eq:30}  
\end{equation}
where $m$ is the dipole moment, $\sigma_s$ is the effective molecular diameter, $\rho^*=\rho\sigma_s^3$ is the reduced density, and $I_n=4\pi\int_0^\infty g_0(r) (dr/r^{n-2})$ is the two-particle perturbation integral calculated based on the pair distribution function $g_0(r)$ of the reference system.\cite{Larsen:77} Correspondingly, $I_{TD}$ is the three-particle perturbation integral involving dipolar interactions between three separate molecular dipoles. More perturbation integrals will enter Eq.\ \eqref{eq:30} when higher molecular multipoles are included in addition to molecular dipoles.\cite{Gubbins:84} 

When the hard-sphere core is used as the reference system, both $I_6(\rho^*)$ and $I_{TD}(\rho^*)$, are functions of $\rho^*$. The inverse temperature $T'/T$ can be therefore viewed as a composite variable $(\rho^*)^{\gamma}/T$, where the density scaling involves the linear factor $\rho^*$ in Eq.\ \eqref{eq:30} and any additional dependence on density from the perturbation integrals.

Similar arguments can be applied to show that $T_K$ increases with increasing pressure.\cite{Roland:2005ga} Specific calculations are, however, harder in this case since they require accounting for the variation of the top of the landscape entropy $\sigma_\infty$ in the parameter $\tau=e_a/\sigma_\infty$ in Eq.\ \eqref{eq:18-1}. For an estimate, one can assume that the shift of $T_K$ comes solely from $T'$. One then gets from Eq.\ \eqref{eq:30} $T_K^{-1} dT_K/dP = \beta_T (1+ \partial \ln [I_{TD}/I_6]/\partial \ln\rho^*)$. For OTP (Table \ref{tab:1}), $T_g^{-1}dT_g/dP\simeq1$ $\mathrm{GPa^{-1}}$,\cite{Roland:2005ga} while the isothermal compressibility is\cite{Naoki:1989uw} $\beta_T\simeq 0.47$ $\mathrm{GPa^{-1}}$. The coefficient in front of the compressibility requires more detailed calculations.

\section{Effect of the electric field}
\label{sec:4}
The external electric field induces a typically weak, anisotropic perturbation of a polar liquid. The corresponding interaction energy adds to the anisotropic interaction energy $H_a$ leading to $H_a(\bm{\mathcal{E}})$. The effect of the external field on the dielectric is nonlocal since the perturbation, $-\sum_j \mathbf{m}_j\cdot \bm{\mathcal{E}}_0$, polarizes all dipoles $\mathbf{m}_j$ in the liquid through the field of external charges  $\bm{\mathcal{E}}_0$.\cite{Hoye:1980gt} The problem is simplified in the mean-field approximation, which replaces the instantaneous field of all dipoles in the liquid with a local cavity field $\bm{\mathcal{E}}_c$ acting on each dipole  
\begin{equation}
  H_a(\bm{\mathcal{E}}) = H_a - \sum_j \mathbf{m}_j\cdot \bm{\mathcal{E}}_c .
  \label{eq:23}
\end{equation}
Here, the cavity field $\bm{\mathcal{E}}_c=\chi_c\bm{\mathcal{E}}$ is connected to the Maxwell field $\bm{\mathcal{E}}$ through the cavity field susceptibility $\chi_c$. It is given as $\chi_c=3\epsilon_s/(2\epsilon_s+1)$ in the dielectric boundary-value problem.\cite{Jackson:99} The new definition of the anisotropic interaction $H_a(\bm{\mathcal{E}})$ can be used in Eqs.\ \eqref{eq:12} and \eqref{eq:12-1} to determine the deformation of the landscape caused by the external field. It turns out that the field affects only $\langle H_a(\bm{\mathcal{E}})^2\rangle$, which becomes
\begin{equation}
  \langle H_a(\bm{\mathcal{E}})^2\rangle = \langle H_a^2\rangle +  (m^2/3)N \mathcal{E}_c^2 . 
  \label{eq:24}
\end{equation}

The second term is in this equation is small compared to the first one at the typical experimental conditions. The smallness parameter is the reduced field $e_c^2 = \mathcal{E}_c^2\sigma_s^6/m^2$, which quantifies the effect of the external field on the molecular-scale interactions between the molecular dipoles. For the Maxwell field $\mathcal{E}\simeq 200$ kV/cm, one gets $e_c^2 \simeq \times 10^{-3}$ at $\sigma_s = 4$ \AA\ and $m=2$ D, making the interaction with the field a small correction to the reduced energy $e_a$ [Eq.\ \eqref{eq:12-1}] in the absence of the field.  

Equation \eqref{eq:24} allows us to calculate the shift of the Kauzmann temperature induced by the field. One starts with Eqs.\ \eqref{eq:12} and \eqref{eq:12-1} establishing the connection between $T'$ and $\tau$, entering the Kauzmann temperature $T_K = T'\left(\sqrt{\tau} -1\right)$, and $\langle H_a(\mathcal{E})^2\rangle$. After some algebra and taking only the main contribution to the temperature change, one obtains
\begin{equation}
k_\text{B}\Delta T_K = \frac{1}{6\sqrt{2}\sigma_\infty}\, \frac{(m\mathcal{E}_c)^2}{k_\text{B}(T'+T_K)} . 
\label{eq:25}  
\end{equation}

The external field thus lowers the entire $s_c(T)$ curve and shifts the Kauzmann temperature to a higher value.\cite{Moynihan:1981ud} According to the AG equation, it makes relaxation slower, in qualitative accord with experiment.\cite{YoungGonzales:2015vp} The parameters $T_K$ and $\sigma_\infty$ are often reported from the analysis of the experimental data.\cite{Richert:98} When added to such data, $\Delta T_K$ provides an estimate of $T'$. This implies that the present energy landscape model can be fully parametrized based on $T_K$, $\sigma_\infty$, and $\Delta T_K$. 

Alternatively, Eq.\ \eqref{eq:25} provides a direct estimate of $\Delta T_K$ when parameters $\sigma_\infty$, $T_K$, and $T'$ are known from fits to excess thermodynamics (Fig.\ \ref{fig:3} and Table \ref{tab:1}). For instance, in the case of  MTHF  ($m=2.1$ D) one gets $\Delta T_K=0.02$ K at $\mathcal{E}=200$ V/cm and $\chi_c=3/2$. A note of caution is relevant here. Our estimate is based on the gas-phase dipole moment $m$. The condensed-phase dipole moment $m'$, enhanced by molecular polarizability,\cite{Boettcher:73,SPH:81} should be used instead in realistic calculations. Since $m'>m$, this correction should lead to a somewhat higher $\Delta T_K$. The value of $m'=2.7$ D for MTHF can be estimated from Wertheim's 1-RPT theory of polarizable liquids\cite{Wertheim:1979bl} yielding $\Delta T_K=0.03$ K.

The dipole moment $m'$ also enters standard mean-field expressions for the dielectric constant of polarizable liquids\cite{SPH:81} and can be alternatively calculated from $\epsilon$ and the high-frequency dielectric constant $\epsilon_\infty$. By neglecting $\epsilon_\infty$ relative to $\epsilon$ and putting $\Delta T_K\simeq \Delta T_g$, one can obtain an estimate of the shift in the glass transition temperature caused by the field
\begin{equation}
k_\text{B}\Delta T_g \simeq \frac{3\epsilon_g}{8\pi\sqrt{2}\rho_g\sigma_\infty}\frac{\epsilon_g-1}{2\epsilon_g+1}\,\frac{\mathcal{E}^2}{1+T'/T_g},
\label{eq:31}  
\end{equation}
where $\epsilon_g=\epsilon(T_g)$ and $\rho_g=\rho(T_g)$. Given that $T'<T_g$ from our results in Table \ref{tab:1}, the above equation can be further simplified at $\epsilon_g\gg 1$ to
\begin{equation}
k_\text{B}\sigma_\infty \Delta T_g \simeq \frac{\epsilon_g\mathcal{E}^2}{8\pi\rho_g} , 
\label{eq:32}  
\end{equation}
where $\epsilon_g\mathcal{E}^2/(8\pi\rho_g)$ is the free energy of the electric field per molecule of the liquid. Equation \eqref{eq:32} has a simple meaning. It suggests that the free energy of the electrostatic field contributes to the shift of the glass transition temperature with the entropy slope given by the top of the landscape entropy $k_\text{B}\sigma_\infty$.  

An alternative estimate of $\Delta T_K$ can be obtained by using the connection between $\langle H_a^2\rangle$ and the perturbation integrals. For $H_a$ representing dipole-dipole interactions the result is $\langle H_a^2\rangle = m^4/(3\sigma_s^6)N\rho^*I_6$. Correspondingly, the shift of the Kauzmann temperature becomes
\begin{equation}
k_\text{B}\Delta T_K = \sigma_s^3 \mathcal{E}_c^2/\sqrt{24\sigma_\infty\rho^*I_6} .
\label{eq:26}  
\end{equation}
For a hard-sphere reference core, the perturbation integral $I_6$ is a function of $\rho^*$, which was tabulated by Larsen \textit{et al}:\cite{Larsen:77} $I_6=4.1888+ 2.8287\rho^*+0.8331(\rho^*)^2 + 0.0317 (\rho^*)^3 + 0.0858(\rho^*)^4-0.0846(\rho^*)^5$.  In the typical range of densities for liquids at 1 atm, $\rho^*\simeq 0.8-0.9$, Eq.\ \eqref{eq:26} gives a crude estimate 
\begin{equation}
k_\text{B}\Delta T_K \simeq 0.1 \sigma_s^3\mathcal{E}_c^2/\sqrt{\sigma_\infty}.   
\label{eq:27}
\end{equation}
The dependence on $m^2$ is canceled out in this approximate equation. The resulting dependence of the thermodynamics on the external field is through the electrostatic energy stored in the volume of the molecule $\sim \sigma_s^3 \mathcal{E}^2$. The cancellation will not occur when $T'$ and $T_K$ are empirical parameters affected by LJ interactions and extracted from the fitting of the excess thermodynamics. The more accurate Eq.\ \eqref{eq:25} should be used instead. Nevertheless, Eq.\ \eqref{eq:27} gives a reasonable estimate of $\Delta T_K$ in the case of MTHF. With $\sigma_\infty=10$ (Table \ref{tab:1}), $\chi_c=3/2$, and $\sigma_s=5.3$ \AA, one gets $\Delta T_K \simeq 0.03$ K at $\mathcal{E}=200$ kV/cm, not far from the above estimate.

\begin{figure}
\includegraphics*[clip=true,trim= 0cm 1cm 0cm 0cm,width=7cm]{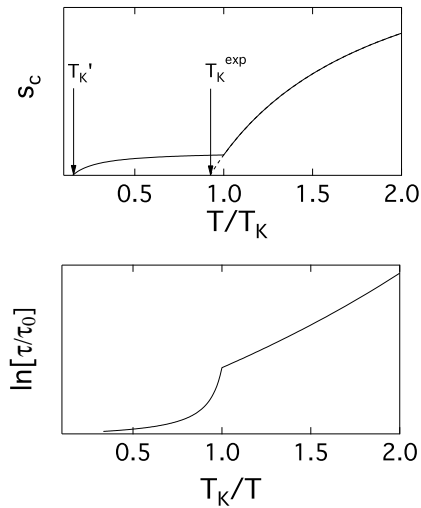}
\caption{Cartoon of the Kauzmann entropy crisis in which the configurational entropy of multipolar interactions vanishes at $T_K$ and the configurational entropy of related to other intermolecular interactions (for instance LJ-type) vanishes at $T_K'$ (upper panel). The overall configurational entropy shows a discontinuous change at the temperature $T_K$. The experimentally reported value $T_K^\text{exp}$ is the result of extrapolating the high-temperature trend to zero. The lower panel shows the corresponding relaxation time in the Arrhenius coordinates calculated from the AG relation.  
}
\label{fig:5}  
\end{figure}

\section{Discussion}
\label{sec:5}
The exact solution for the enumeration function presented here allows both scenarios, with an ideal glass and its avoidance. It is important to recognize that the state of zero configurational entropy is reached in the ideal glass scenario for orientational degrees of freedom only since these are the motions predominantly affecting multipolar interactions. A small residual configurational entropy arising from translations altering the local molecular packing can still exist. This result might carry general significance since it allows one to think of the Kauzmann entropy crisis, originating from extrapolating the excess entropy to zero line, as the consequence of the entropy drop from a subset of the liquid degrees of freedom. The low-temperature liquid will still possess a non-vanishing configurational entropy, which might decay to zero at a separate Kauzmann temperature $T_K'$. The overall decay of the configurational entropy as temperature is reduced might look as sketched in the upper panel of Fig.\ \ref{fig:5}. When translated to relaxation dynamics by using the AG relation, vanishing orientational entropy leads to a dynamic crossover of fragile to strong type\cite{Ito:99} (lower panel in Fig.\ \ref{fig:5}). The temperature of dynamical crossover will generally be higher than the experimentally reported Kauzmann temperature $T_K^\text{exp}$ produced by extrapolating the high-temperature entropy to zero.   

Whether the low-temperature portion of the configurational entropy will show a significant change with temperature depends on the glass former. Many glass formers have their glass and crystalline heat capacities very close below $T_g$.\cite{Tatsumi:2012iy} In a number of other cases, such as monoalcohols\cite{Kabtoul:2008ie} and toluene,\cite{Yamamuro:1998td} the heat capacity of the glass just below $T_g$ is above that of the crystal and then merges with crystal's heat capacity with lowering temperature. This latter case would correspond to a noticeable temperature variation of the low-temperature entropy in Fig.\ \ref{fig:5}. Still, even in the case of monoalcohols, molecular rotations are responsible for the main part of $s_\text{ex}$. This is demonstrated by close values of heat capacities of supercooled ethanol and its plastic crystal phase.\cite{Kabtoul:2008ie,Ramos:2013wp}  

\begin{figure}
\includegraphics*[clip=true,trim= 0cm 1cm 0cm 0cm,width=7cm]{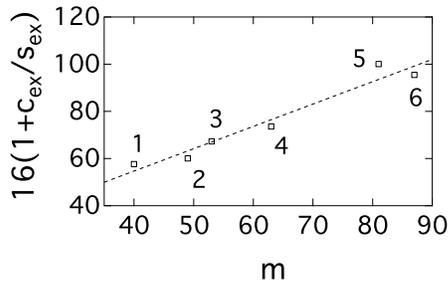}
\caption{{\label{fig:6}}$16\left(1 + c_\text{ex}/s_\text{ex}\right)$ [Eq.\ \eqref{eq:40}] taken from Ref.\ \onlinecite{KleinAngell:16} vs kinetic fragility\cite{Wang:2006cb,Wang:2010df} for a number of liquid glass formers: propanol (1), 1,3-propandiol (2),\cite{Wang:2010df} glycerol (3), 1-butene (4),  $o-$terphenyl (5),    1,3-diphenyl-1,1,3,3-tetramethyldisiloxane (6). The points are obtained by extrapolating $c_\text{ex}(T)/s_\text{ex}(T)$ from Ref.\ \onlinecite{KleinAngell:16} to $T=T_g$. The dashed line is the linear regression with the slope 0.95. }    
\end{figure}

The scenario of two entropy components, with a nearly temperature-independent low-temperature part, can be connected to relaxation data within the AG scheme. Since the experimental Kauzmann temperature $T_K^\text{exp}$ is below $T_K$, one can assume $T_K\simeq T_g$ and write the excess entropy in the form $s_\text{ex}(T)=s_d + c_\text{ex}(1-T_K/T)$, where $s_d$ is the entropy component in excess to the orientational entropy (mostly from thermal agitation of the density). With this form, one gets for the liquid kinetic fragility\cite{Angell:95,Wang:2006cb} 
\begin{equation}
 m = \frac{d\log \tau }{d(T_g/T)}=16\left(1 + c_\text{ex}/s_\text{ex}\right), 
 \label{eq:40}
\end{equation}
where, as above,  $c_\text{ex}$ and $s_\text{ex}$ are measured at $T_g$. The connection of fragility to the ratio $c_\text{ex}/s_\text{ex}$ was recently recognized by Klein and Angell.\cite{KleinAngell:16}  Their compilation of data is consistent with Eq.\ \eqref{eq:40} (Fig.\ \ref{fig:6}). 

It is often stated that the ideal glass state is not reachable because a macroscopic system will always possess thermal excitations at a positive temperature.\cite{Mauro:2014df} While this statement is generally correct, it misses the point that the corresponding configurational entropy will be zero, in the thermodynamic limit, if such thermal excitation produce subexponential enumeration with respect to the number of molecules $N$.\cite{Stillinger:2013tl} Our derivation of the enumeration function performed for a finite $N$ [Eq.\ \eqref{eq:13-1}] clearly demonstrates this point. The transition, in the thermodynamic limit $N\to\infty$, from Eq.\ \eqref{eq:13-1} to the enumeration function in Eq.\ \eqref{eq:14} involves neglecting the subexponential terms in the density of states scaling as $(2N)^{-1}\ln N$. Other examples with subexponential scaling might include excitations at the grain boundaries of well-packed regions\cite{Langer:06} producing heterogeneous structure in the low-temperature liquid. All such excitations, while present, will not contribute to the enumeration function calculated in the thermodynamic limit. In terms of the two-entropy picture shown in Fig.\ \ref{fig:5}, the orientational excitations will enumerate subexponentially below $T_K$, while density excitations will enumerate exponentially.

The present model shows that the configurational entropy arising from anisotropic multipolar interactions is decreased by the electric field. One has to realize that the model produces a nonlinear effect of the field on the liquid structure. Mathematically, this is easy to realize by noting that both effective temperatures $\tau(\mathcal{E}^2)$ and $T'(\mathcal{E}^2)$ in Eq.\ \eqref{eq:17} for the configurational entropy are nonlinear functions of $\mathcal{E}^2$. When expanded in series of $\mathcal{E}^2$, the configurational entropy and the Kauzmann temperature scale linearly with $\mathcal{E}^2$ in the lowest expansion term of main interest for experiment.  

From the general perspective, a non-linear effect of the electric field on the liquid can modify its structure and change its relaxation time. This result is opposite to what is expected from linear response, which preserves the structure and relaxation of the unperturbed liquid. The linear response is in fact assumed\cite{Landau8} in deriving the thermodynamics of a polarized liquid in Eq.\ \eqref{eq:3}. From this general argument, it seems impossible for such linear polarization to modify the relaxation dynamics.  

The distinction between the present nonlinear model and Eq.\ \eqref{eq:3} can be further appreciated by looking at the variance of the anisotropic interaction energy in Eq.\ \eqref{eq:24}, which eventually defines $\tau(\mathcal{E}^2)$ and $T'(\mathcal{E}^2)$. It shows that the field term in the variance of  $H_a(\bm{\mathcal{E}})$ involves only the one-particle orientational fluctuations of separate liquid dipoles and does not involve correlations between dipolar rotations (of binary or higher order type). In contrast, the temperature derivative of the dielectric constant in the thermodynamic entropy in Eq.\ \eqref{eq:3} is determined by higher-order, triple and four-particle, correlations between the dipoles.\cite{DMjcp1:16} It is therefore hard to see how the use of the thermodynamic polarization entropy to alter the configurational entropy can be reconciled with the present microscopic model.

\acknowledgments 
This research was supported by the National Science Foundation (CHE-1464810). Discussions with Ranko Richert and Austen Angell are gratefully acknowledged. 

%\bibliography{chem_abbr,dielectric,dm,statmech,protein,liquids,solvation,dynamics,simulations,surface,water,glass,nano,lcold,ir}


\end{document}